\documentclass[onefignum,onetabnum]{siamart171218}

\usepackage{lipsum}
\usepackage{amsfonts}
\usepackage{graphicx}
\usepackage{epstopdf}
\usepackage{algorithmic}
\usepackage{mathtools}
\ifpdf
  \DeclareGraphicsExtensions{.eps,.pdf,.png,.jpg}
\else
  \DeclareGraphicsExtensions{.eps}
\fi

\newsiamremark{remark}{Remark}
\newsiamremark{hypothesis}{Hypothesis}
\crefname{hypothesis}{Hypothesis}{Hypotheses}
\newsiamthm{claim}{Claim}

\headers{Optimization of Constrained QC Mapping for Origami Design}{K. H. Lai, H. T. Tsang, G. P. T. Choi, L. M. Lui}

\title{Optimization of Constrained Quasiconformal Mapping for Origami Design\thanks{Submitted to the editors DATE.
\funding{TODO.}
}
}

\author{
Ka Ho Lai\thanks{Department of Mathematics, The Chinese University of Hong Kong
  (\email{khlai@math.cuhk.edu.hk}).}
\and Hei Tung TSANG\thanks{Department of Mathematics, The Chinese University of Hong Kong \\ (\email{httsang@math.cuhk.edu.hk}).}
\and Gary P. T. Choi\thanks{Department of Mathematics, The Chinese University of Hong Kong
  (\email{ptchoi@cuhk.edu.hk}).}
\and Lok Ming Lui\thanks{Department of Mathematics, The Chinese University of Hong Kong
  (\email{lmlui@math.cuhk.edu.hk}).}
}

\usepackage{amsopn}

\usepackage{multirow}

\makeatletter
\newcommand*{\addFileDependency}[1]{
  \typeout{(#1)}
  \@addtofilelist{#1}
  \IfFileExists{#1}{}{\typeout{No file #1.}}
}
\makeatother

\ifpdf  
\hypersetup{
    pdftitle={
        Optimization of Constrained Quasiconformal Mapping for Origami Design
    },
    pdfauthor={
        K. H. Lai, H. T. Tsang, G. P. T. Choi and L. M. Lui
    }
}
\fi

\begin{document}

\maketitle

\begin{abstract}
Origami structures, particularly Miura-ori patterns, offer unique capabilities for surface approximation and deployable designs. In this study, a constrained mapping optimization algorithm is designed for designing surface-aligned Miura-ori via a narrow band approximation of the input surface. The Miura-fold, embedded in the narrow band, is parameterized to a planar domain, and a mapping is computed on the parameter pattern by optimizing certain energy terms and constraints. Extensive experiments are conducted, showing the significance and flexibility of our methods. 
\end{abstract}


\section{Introduction}
\label{sec: introduction}
Origami is the Japanese word for ``folding paper". It is the art to turn a flat sheet into various folded structures. Miura-ori is a special form of origami, whose crease pattern forms a tessellation of quadrilaterals. Its unique geometric structure provides interesting physical properties. Miura-ori folds allow rigid materials to deform while maintaining their structure. Moreover, without changing the material itself, it provides a lightweight, yet structurally stiff form. It has been widely applied in various areas, including designing mechanical metamaterials \cite{lv2014origami, schenk2013geometry, zhou2016origami}, deployable structures \cite{filipov2015origami, koryo1985method, wang2023design}, and biomedical devices \cite{johnson2017fabricating, kuribayashi2006self}. In recent years, many studies on designing surface-aligned Miura-ori patterns have been conducted \cite{dang2022inverse, dudte2021additive, dudte2016programming, hu2021constructing, wang2016folding}, as a generalization of conventional flat-shaped Miura-ori folds. Miura-ori folds with curvature provides flexibility to designing more general deployable structures and devices. 

In this paper, we formulate the problem of designing surface-aligned Miura-ori patterns as a constrained mapping optimization problem. For a given surface, its upper and lower surfaces are constructed using the normal vectors, creating a narrow band approximation of the original surface. The folded pattern is embedded between the upper and lower surfaces and parameterized into a planar domain, and optimized under specific shape regularizers and unfoldability constraints. In particular, quasiconformality is included as a regularization to deformation in the parameter domain. As a generalization of conformal mappings, quasiconformal mappings offer flexibility by controlling distortion. Using the Beltrami equation, the distortion can be measured by the Beltrami coefficient $\mu$. Combining the other two regularization terms for size and position, the optimization is performed under flatness and developability constraints, which are crucial for achieving unfoldability. Experiments are conducted in various manners to examine the significance of regularizers, the flexibility on various surfaces, and the impact of thickness and resolution on the optimization results.

\section{Related Works}

The mathematical and computational study of origami has led to significant advances in understanding its geometry, mechanics, and design principles \cite{demaine2007geometric, dudte2021additive, hull1994mathematics, lang1996computational, lang2012origami}. Among the various origami patterns, the Miura-ori \cite{miura1985method} stands out due to its simplicity, versatility, and unique mechanical properties. The Miura-ori pattern has been extensively studied for its applications in deployable structures, metamaterials, and engineering design. Several direct geometrical methods have been developed for designing the Miura-ori pattern \cite{feng2020helical, gattas2013miura, gattas2014miura, hu2019design, mitani2009design, sareh2015design, song2017design, wang2016folding, zhou2015design}, while optimization techniques have been used to ensure specific properties such as foldability, minimal distortion, and structural stability \cite{dudte2016programming, jiang2019curve, kilian2008curved, tachi2009generalization}. The mechanical properties \cite{schenk2013geometry, silverberg2014using, silverberg2015origami, waitukaitis2015origami, wei2013geometric} and geometric characteristics \cite{gattas2014miura, tachi2009generalization} of the Miura-ori pattern have also been extensively analyzed.

The constrained mapping problem has been studied extensively in the context of geometric optimization. Numerous approaches have focused on optimizing diffeomorphic mappings, where the mapping is required to be smooth and bijective \cite{lui2015splitting, zeng2014surface}. Quasiconformal mappings have frequently been employed in these studies due to their ability to handle controlled distortions while preserving the essential structure of the mapping \cite{lui2013texture}. In addition, specific mapping problems often necessitate landmark registration \cite{lam2014landmark, lyu2024bijective, zhang2022unifying} or intensity registration \cite{lam2014landmark, zhang2022unifying}, where sets of corresponding points or regions must align precisely. These approaches demonstrate the importance of incorporating constraints into mapping problems to meet practical requirements.

\section{Mathematics Background}
\label{sec: math background}
\subsection{Quasiconformal maps}
\label{sec: qc}
Quasiconformal maps, a generalization of conformal maps, are briefly described here. 

\begin{definition}
    \label{def: quasiconformal map}
    An complex valued function $f:\Omega \subseteq \mathbb{C} \to \mathbb{C}$ is called quasiconformal if it satisfies the Beltrami equation
    \begin{equation}
    \label{eqn: beltrami}
        \frac{
            \partial f
        }{
            \partial \Bar{z}
        } 
        = 
        \mu (z)
        \frac{
            \partial f
        }{
            \partial z
        },
    \end{equation}
    for some complex-valued Lebesgue measurable function 
    $\mu$ 
    with $\| \mu \|_{\infty} < 1$. 
\end{definition}
The \textit{Wirtinger derivatives} are used in the above definition, which are defined as
\begin{equation}
    \frac{
        \partial 
    }{
        \partial z
    } = 
    \frac{1}{2}
    \left(
    \frac{
        \partial
    }{
        \partial x
    }
    -
    i \frac{
        \partial
    }{
        \partial y
    }
    \right)
    \quad \text{and} \quad
    \frac{
        \partial 
    }{
        \partial \Bar{z}
    } = 
    \frac{1}{2}
    \left(
    \frac{
        \partial
    }{
        \partial x
    }
    +
    i \frac{
        \partial
    }{
        \partial y
    }
    \right) .
\end{equation}

Moreover, the Jacobian determinant of the function $f$ is 
\begin{equation}
    \det J_f = | f_z |^2 ( 1 - | \mu_f |^2 ). 
\end{equation}
As a result, if $\| \mu \|_\infty < 1$, then $\det J_f > 0$ and hence $f$ is bijective. 

If $\mu \equiv 0$, then $f$ is a planar conformal map. The \textit{Beltrami coefficient} $\mu$ can be used as a measure of the conformality of $f$. If $f$ is quasiconformal, then the \textit{maximal quasiconformal dilation} of $f$ is given by 
\begin{equation}
    K = \frac{1 + \| \mu \|_\infty}{1 - \| \mu \|_\infty}.
\end{equation}

\begin{figure}[t]
    \centering
    \includegraphics[width=.7\textwidth]{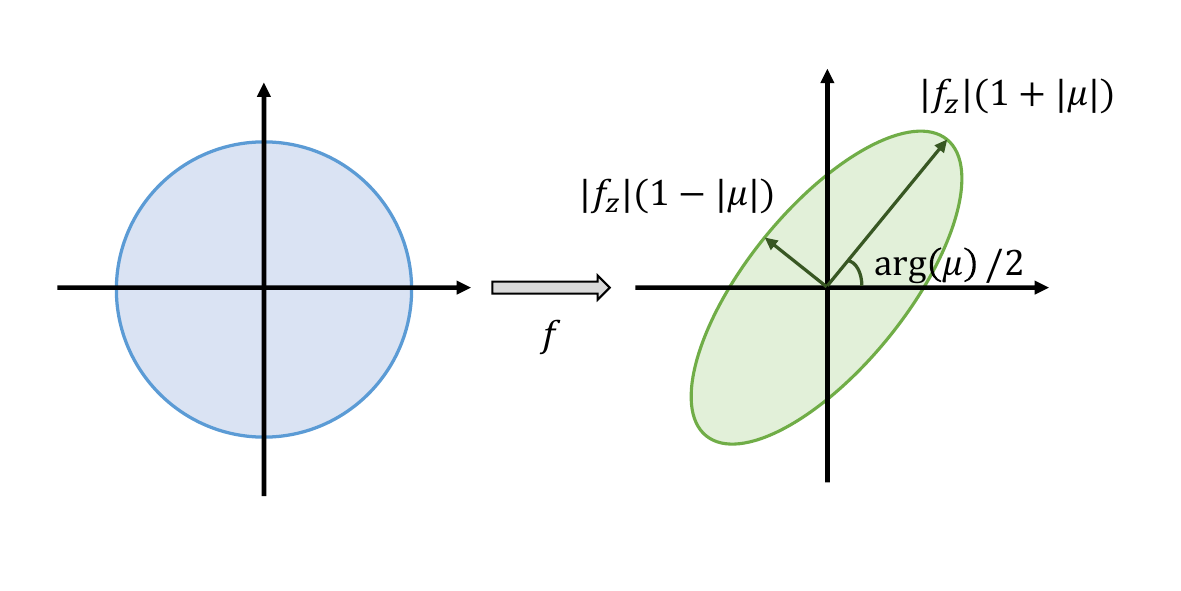}
    \caption{
        An illustration of quasiconformality. 
    }
    \label{fig: qc}
\end{figure}

This suggests that we can control angle distortion by controlling the magnitude of $\mu$. 
The idea of quasiconformality is illustrated in Figure \ref{fig: qc}.

\subsection{Necessary Conditions for Miura-Ori Pattern}
Miura-ori is a classic origami tessellation widely studied for its applications in engineering, design, and deployable structures. In \cite{dudte2016programming}, Dudte et al. formulate two mathematical constraints that a Miura-ori pattern must satisfy to ensure both physical feasibility.

To describe the pattern mathematically, it is modeled as a quadrilateral mesh structure $(V, Q)$, where $V$ represents the set of nodes or vertices, and $Q$ denotes the quad mesh connectivity, specifying how the vertices are connected to form quadrilateral faces. This representation provides a convenient framework for analyzing and enforcing the constraints.

\subsubsection*{Planarity Constraint}

The first constraint is the \textit{planarity constraint}, which ensures that each quadrilateral face in the folded pattern remains flat. This is critical because, in a valid folded state, the faces of the origami structure must lie in a plane, with no curvature or bending within the face. Mathematically, this means that the four vertices of each quadrilateral face must be coplanar, resulting in a face with zero volume. For a given quadrilateral face $q \in Q$, with vertices labeled as $a_q$, $b_q$, $c_q$, and $d_q$, the planarity condition can be expressed as:
\begin{equation}
    0 = g_{\text{planarity}}(q) = [(b_q - a_q) \times (c_q - a_q)] \cdot (d_q - a_q),
\end{equation}
where the cross product $(b_q - a_q) \times (c_q - a_q)$ defines the normal vector of the plane formed by the vertices $a_q$, $b_q$, and $c_q$, and the dot product with $(d_q - a_q)$ ensures that the fourth vertex $d_q$ lies in the same plane. This condition must be satisfied for all faces in the mesh.

\subsubsection*{Developability Constraint}

The second constraint is the \textit{developability constraint}, which ensures that the folded pattern can be flattened into a two-dimensional plane without tearing or overlapping. In origami terms, this means that the sum of the angles around each interior vertex must equal $2\pi$, reflecting the geometric property of a flat, undeformed sheet. 

For an interior vertex $v \in V$, let $\{ \theta^i_v \}_{i=1}^m$ denote the set of angles around $v$, where $m$ is the number of adjacent quadrilateral faces meeting at $v$. The developability condition is expressed as:
\begin{equation}
    0 = g_{\text{develop}}(v) = 2\pi - \sum_{i=1}^m \theta^i_v.
\end{equation}
This constraint ensures that no excess material accumulates around any vertex, preserving the ability of the origami pattern to transition smoothly between folded and flat states.

\subsubsection*{Combined Constraints for the Quadrilateral Mesh}

To simplify notation, the two constraints can be extended to the entire quadrilateral mesh $(V, Q)$ as follows:
\begin{equation}
    g_{\text{planarity}}(V, Q) = 0
\end{equation}
and
\begin{equation}
    g_{\text{develop}}(V, Q) = 0.
\end{equation}

These equations represent the global conditions that the mesh must satisfy to be a valid Miura-ori pattern. The planarity constraint, applied to each quadrilateral face, guarantees that all faces are flat, while the developability constraint, applied to each interior vertex, ensures that the folded structure remains geometrically consistent.

Together, these planarity and developability constraints form the foundation for designing and analyzing Miura-ori patterns. They enable researchers and engineers to explore a wide range of applications, from deployable solar panels and foldable architecture to metamaterials with tunable mechanical properties. Furthermore, these mathematical formulations have inspired computational methods for generating complex origami patterns and studying their mechanical behavior under various loading conditions. By satisfying these constraints, the Miura-ori pattern achieves its remarkable combination of flexibility, compactness, and structural stability.

\section{Proposed Methods}

\subsection{Optimization Model}
\label{sec: optimization model}
For a given surface, the objective is to determine a Miura-ori pattern that closely approximates the geometry of the surface. In this section, this inverse design problem is formulated as a constrained mapping problem, characterized by complex and nonlinear constraints. The process involves constructing a mapping that not only satisfies these constraints but also preserves important geometric and structural properties required for the origami design. 

Let $\mathcal{M}$ be the target surface for the origami pattern. Suppose we parameterized it with $\phi: R \subset \mathbb{R}^2 \to \mathcal{M}$, where $R$ is a simply-connected bounded domain. The parameterization $\phi$ provides a way to map points from the 2D domain $R$ to the 3D surface $\mathcal{M}$. If $\mathcal{M}$ is smooth, we can compute the normal vectors $\Vec{n}: R \to \mathbb{R}^3 $ of $\mathcal{M}$. These normal vectors are critical for defining surfaces parallel to $\mathcal{M}$. Using these, we can define an upper surface $\mathcal{M}^{u}$ and a lower surface $\mathcal{M}^l$ of $\mathcal{M}$, which are parameterized by 
\begin{equation}
\begin{split}
    \phi^u(x, y) &= \phi(x, y) + \epsilon \Vec{n}(x, y), \\
    \phi^l(x, y) &= \phi(x, y) - \epsilon \Vec{n}(x, y),
\end{split}
\end{equation}
where $\epsilon > 0$ is a constant. These surfaces are offset above and below the target surface $\mathcal{M}$ by a small distance $\epsilon$ and will serve as the bounding surfaces of the folded origami structure. This offset ensures that the final folded pattern remains within a prescribed maximum deviation from the original surface $\mathcal{M}$. 

In the domain $R$, a quadrilateral mesh $(V_0, Q)$ is constructed, where $V_0 \subset R$ is the set of vertices, and $Q$ denotes the quadrilateral connectivity. The vertices $V_0$ are partitioned into two sets, $V^u_0$ and $V^l_0$, corresponding to the upper and lower surfaces, respectively. The initial folded pattern is denoted as $P_0 = \phi^u(V^u_0) \bigcup \phi^l(V^l_0)$. The detailed procedure for constructing this initial folded pattern is discussed later in Subsection \ref{sec: initial guess}. 

In the optimization process, the planar pattern $(V_0, Q)$ is deformed to $(f(V_0), Q)$, where $f: R \to R$ is the planar mapping to be optimized. The corresponding folded pattern is $P(f) = \phi^u(f(V^u_0)) \bigcup \phi^l(f(V^l_0))$. The goal of the optimization is to determine a planar mapping $f^*$ such that the resulting folded pattern satisfies the required constraints and approximates the target surface $\mathcal{M}$. Mathematically, this optimization problem is expressed as:
\begin{equation}
\begin{split}
    f^* &= \text{argmin}_{f} E(f) 
    \\ 
    \text{subject to } \quad 0 &= g_{\text{planarity}}(P(f), Q) 
    \\
    0 &= g_{\text{develop}}(P(f), Q). 
\end{split}
\end{equation}

In this formulation, $g_{\text{planarity}}$ ensures that the quadrilaterals in the folded pattern remain planar, which is a critical requirement for physical foldability. The constraint $g_{\text{develop}}$ ensures the developability of the folded pattern, meaning that the pattern can be constructed from a flat sheet without stretching or tearing. 

To guide the optimization process, an energy function $E(f)$ is defined, consisting of multiple terms that regularize different aspects of the deformation. The first component of this energy is the edge length energy, which measures the deviation of edge lengths in the folded pattern from their initial values. Denote $L^i_0$ as the length of edges in $(P_0, Q)$, and $L^i_f$ as the length of edges in $(P(f), Q)$. Following the idea in~\cite{dudte2016programming}, the edge length energy is defined as:
\begin{equation}
\label{ene: edge length}
\begin{split}
    E_l (f) = \frac{1}{|L|} \sum_{i} \frac{(L^i_f - L^i_0)^2}{2 L^i_0}, 
\end{split}
\end{equation}
where $|L|$ is the total number of edges for normalizing the edge length energy. This term penalizes changes in edge lengths, ensuring that the folded pattern remains consistent with the initial design. By minimizing $E_l(f)$, the optimization process preserves the geometric proportions of the folded origami structure. 

To further regularize the deformation, an energy on quasiconformality is introduced. Quasiconformality provides a measurement of how much a mapping deviates from being angle-preserving. Each quadrilateral face in $Q$ is divided into two triangular faces, forming a triangular mesh structure $T$. Discrete Beltrami coefficients, which quantify the local distortion of the mapping, can be computed for each triangular face in the initial pattern $(V_0, T)$ and their deformation $(f(V_0), T)$, following the approach in \cite{lui2013texture}. The quasiconformality energy is then defined as:
\begin{equation}
\label{ene: beltrami}
\begin{split}
    E_{\mu}(f) = \frac{1}{|T|} \sum_{t \in T} | \mu_t(f) |^2,
\end{split}
\end{equation}
where $|T|$ is the total number of triangular faces. Minimizing $E_{\mu}(f)$ ensures that the planar mapping $f$ preserves the local geometry of the pattern as much as possible, reducing distortions in angles and shapes during deformation. This energy plays a crucial role in maintaining the structural feasibility of the folded pattern. 

Additionally, to address potential issues with translation, a centering energy is introduced. Empirically, it has been observed that the optimized pattern can sometimes shift away from the desired position, resulting in an undesired final pattern. The centering energy is defined as:
\begin{equation}
\label{ene: centering}
\begin{split}
    E_c (f) = \frac{1}{|V|} \sum_{(v_x, v_y) \in f(V)} \left(\left(\frac{v_x - c_x}{R_x}\right)^2 + \left(\frac{v_y - c_y}{R_y}\right)^2\right), 
\end{split}
\end{equation}
where $c = (c_x, c_y)$ is a pre-defined point, typically chosen as the center of the region of interest, $R_x, R_y$ are the range of the domain $R$ in $x$ and $y$ respectively, and $|V|$ is the total number of vertices. This energy encourages the optimized pattern to remain aligned with the initial configuration. Although the weight of this energy is kept small to avoid dominating the optimization process, it is essential for preventing significant translations or shifts of the pattern. 

The final optimization model is expressed as:
\begin{equation}
\label{eqn: final optimization model}
\begin{split}
    f^* &= \text{argmin}_{f} \rho_1 E_l (f) + \rho_2 E_\mu (f) + \rho_3 E_c (f)
    \\ 
    \text{subject to } \quad 0 &= g_{\text{planarity}}(P(f), Q) 
    \\
    0 &= g_{\text{develop}}(P(f), Q), 
\end{split}
\end{equation}
where $\rho_1, \rho_2, \rho_3 \geq 0$ are weights that balance the contributions of the different energy terms. $E_l$ ensures consistency in edge lengths, $E_\mu$ minimizes shape distortions, and $E_c$ maintains the position of the pattern. 

This formulation ensures that the final folded pattern approximates the target surface $\mathcal{M}$ within a maximum deviation of $\epsilon$. The combination of constraints and energies provides a robust framework for designing Miura-ori patterns that are both geometrically accurate and physically feasible. By balancing these components, the optimization process generates patterns that closely conform to $\mathcal{M}$ while preserving the desired structural and geometric properties of the origami design.

\subsection{Initial Pattern Construction}
\label{sec: initial guess}

Here, the construction details of the initial guess are discussed. 

\begin{figure}[htbp]
    \centering
    \includegraphics[width=\textwidth]{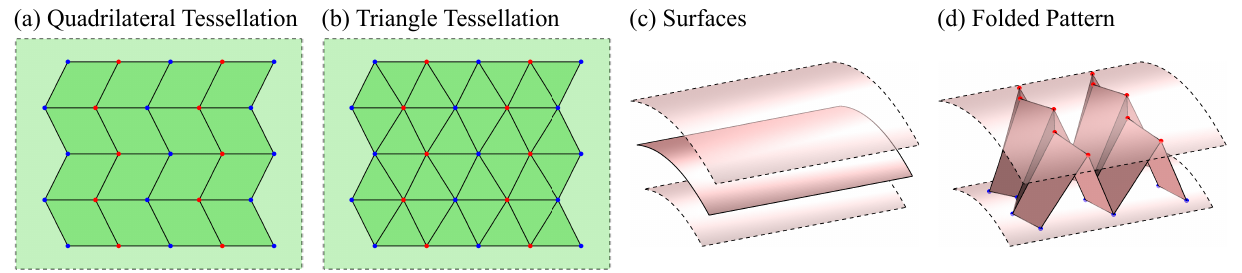}
    \caption{
        Construction of the initial patterns.
        (a) and (b) illustrate the parameter domain and initial parameter pattern under quadrilateral and triangular connectivity. (c) shows the input surface together with its upper and lower surfaces. (d) demonstrates the initial folded pattern embedded between the upper and lower surfaces. 
        The red points form $V_u$, the blue points form $V_l$. 
    }
    \label{fig: initial patterns}
\end{figure}

Initially, a tessellation of identical parallelograms is formed in the parameter domain. This process can be seen as constructing a regular grid, and then skewing every two rows horizontally by the same amount. We identify the nodes in the odd columns as lower nodes, forming $V_l$, and $V_u$ is formed by the rest of the points. This construction is illustrated in Figure \ref{fig: initial patterns}(a), with blue and red points indicating $V_u$ and $V_l$ respectively. 

In the optimization model mentioned in the previous Section \ref{sec: optimization model}, discretized Beltrami coefficients are computed in the optimization objective. Following the approach in \cite{lui2013texture}, a triangular mesh is required in computing the discretized Beltrami coefficients. Therefore, each quadrilateral is divided in a Delaunay manner into two triangles. The resulting triangular mesh is illustrated in Figure \ref{fig: initial patterns}(b). 

Given the constructed parameter pattern and the partition of $V$, the construction of the initial folded pattern is simple. The points in $V_l$, which are the blue points in the figures, are mapped to the lower surface, and $V_u$, the red points, are mapped to the upper surface, forming an initial folded pattern. Figure \ref{fig: initial patterns}(c) illustrates the input surface $M$, where the two transparent surfaces are the upper and lower surfaces. In Figure \ref{fig: initial patterns}(d), $V_u$ and $V_l$ are mapped to the corresponding positions by $\phi^u$ and $\phi^l$. 

\section{Numerical Implementation}
With the discussed energy functionals and unfoldability constraints, some implementation details, including the computation of the discretized Beltrami coefficient, and the optimization algorithm, are described here. 

\subsection{Computation of Quasiconformality Energy}
Computing the Beltrami coefficient $\mu(f)$ is the crucial part of the quasiconformality energy. We follow the approach proposed by \cite{lui2013texture}. We first compute the Beltrami coefficient $\mu(f)$, where $f$ is a linear homeomorphism mapping the initial pattern $(V_0, T)$ to its deformation $(f(V_0),T)$ in a discrete sense. As $\mu(f) = \frac{\partial f}{\partial \bar{z}}/ \frac{\partial f}{\partial z} $, the partial derivatives at each $t\in T$ is required to be approximated. Since $f$ is linear, the restriction of $f$ on each triangular face $t \in T$ can be denoted as: 
\[f|_t(x,y) = \left( \begin{array}{cc}
     a_tx + b_ty + r_t  \\
     c_tx + d_ty + q_t
\end{array}\right).\]
Hence, the partial derivatives of $f|_t$ are $\partial_xf(t) = a_t + ic_t$ and $\partial_yf(t) = b_t + id_t$. The gradient $\nabla_t f $ is defined as 
\[ \nabla_t f \coloneq \left( \begin{array}{cc}
    \partial_xf(t)\\
    \partial_yf(t)
\end{array}\right) .\]
Let $(\alpha_t, \beta_t, \gamma_t)$ be the coordinates of the three vertices of triangular face $t \in T$, $e_1 = \beta_t - \alpha_t$ and $e_2 = \gamma_t - \alpha_t$. Then, $\nabla_t f$ can be found by solving the linear equations: 

\[\left( \begin{array}{cc}
    e_1\\
    e_2
\end{array}\right) \nabla_t f = \left( \begin{array}{cc}
    f_t(\beta_t) - f_t(\alpha_t)\\
    f_t(\gamma_t) - f_t(\alpha_t)
\end{array}\right).\]
Then, $a_t$, $b_t$, $c_t$ and $d_t$ can be solved from the above equation, and the Beltrami coefficient $\mu_t(f)$ of each triangular face $t \in T$ can be computed by 
\begin{equation}
    \mu_t(f) = \frac{(a_t - d_t) + i(c_t+b_t)}{(a_t+d_t) + i(c_t-b_t)}.
\end{equation}
With the computed discretized Beltrami coefficients, the quasiconformal energy from Eq.~\eqref{ene: beltrami} can be obtained: 
\[E_\mu (f) = \frac{1}{|T|} \sum_{t \in T} | \mu_t(f) |^2.\]

\subsection{Optimization Method}
In this section, an iterative method used for solving the prescribed optimization problem is discussed. 

In the optimization problem \eqref{eqn: final optimization model}, $f$ is represented by $f(V)$, which is a set of point-wise coordinates. For simplicity, denote $f(V) = y$, and rewrite the optimization problem as 
\begin{equation}
\label{eqn: simplify optimization model}
\begin{split}
    y^* &= \text{argmin}_{y} E(y) \\
    \text{subject to } \quad 0 &= g(y),
\end{split}
\end{equation}
where $g$ is an $n$-dimensional constraint function, and $n$ is the number of constraints, which equals the total number of quadrilateral faces and interior nodes. These constraints enforce the planarity and developability conditions required for a valid Miura-ori pattern. The objective function $E(y)$ includes regularization terms that promote desirable properties, such as geometric smoothness or minimal distortion.

The Lagrangian of the optimization problem~\eqref{eqn: simplify optimization model} is defined as:
\begin{equation}
\label{eqn: lagrangian}
\begin{split}
    L(y, \lambda) = E(y) + \lambda^T g(y), 
\end{split}
\end{equation}
where $\lambda$ is the Lagrange multiplier. The Lagrangian formulation combines the objective function and the constraints, allowing the problem to be analyzed and solved within a unified framework. This formulation is particularly suitable for constrained optimization problems, such as the Miura-ori inverse design, where the constraints are nonlinear and must be satisfied exactly.

Write $J$ to denote the Jacobian of $g$, and $H$ to denote the Hessian of the Lagrangian \eqref{eqn: lagrangian} with respect to $y$, 
\begin{equation}
\begin{split}
    H = \nabla^2 E(y) + \sum_i \nabla^2 \lambda_i g_i(y). 
\end{split}
\end{equation}
The Jacobian $J$ represents the first-order derivatives of the constraints $g(y)$, while the Hessian $H$ captures the second-order derivatives of the objective function and constraints. Together, these matrices provide the necessary information to efficiently compute updates for the iterative method.

In order to solve $\nabla L = 0$, Newton's method is employed. Newton's method is particularly effective for solving nonlinear optimization problems due to its quadratic convergence near the solution. In each iteration, the following linear system must be solved:
\begin{equation}
\label{eqn: Newton}
\begin{split}
    \begin{bmatrix}
        H & J^T \\
        J & 0
    \end{bmatrix}
    \begin{bmatrix}
        \Delta y \\
        \Delta \lambda
    \end{bmatrix}
    =
    \begin{bmatrix}
        \nabla E(y) + J(y)^T \lambda \\
        g
    \end{bmatrix}, 
\end{split}
\end{equation}
where $\Delta y$ and $\Delta \lambda$ represent the updates to the variables $y$ and the Lagrange multipliers $\lambda$, respectively. Once the linear system is solved, the variables are updated as:
\begin{equation}
\begin{split}
    y &\leftarrow y + \Delta y, \\
    \lambda &\leftarrow \lambda + \Delta \lambda.  
\end{split}
\end{equation}
The iteration continues until all the values in $g(y)$ are smaller than the feasibility error tolerance, and a stationary point of $L$ is reached. This ensures that the solution satisfies the constraints while minimizing the objective function. The outline of the optimization scheme is shown in Algorithm \ref{alg: optimization}. 

\begin{algorithm}
    \caption{Optimization Scheme}
    \label{alg: optimization}
    \begin{algorithmic}[1]
        \item[]
        \textbf{Input:}
            Surface Parameterization function $\phi$; 
            Energy weightings: $\rho_1$, $\rho_2$ and $\rho_3$;
            Energy functions: $E_l$, $E_\mu$ and $E_c$;
            Constraint function: $g$; 
            Structure parameters: dimension of Miura-ori pattern $(m,n)$, thickness $\epsilon$, error tolerance: $\sigma_1$, $\sigma_2$;  
            Number of maximum iterations $N$. 
        \item[]
        \textbf{Output:}
            Optimized parameter pattern $y$ 
        \STATE
            Generate an initial pattern $(V_0,Q)$ using $(m,n)$;
        \STATE
            Compute normal vectors and form $\phi^u$ and $\phi^l$ using $\epsilon$; 
        \STATE
            $y \gets V_0$;
        \FOR{$i = 1, \dots, N$}
        \STATE
            Compute the folded pattern $P = \phi^u(y) \bigcup \phi^l(y)$;
        \STATE
            Compute the Hessian $\nabla^2E = \nabla^2(\rho_1E_l + \rho_2E_\mu + \rho_3E_c)$ using $V_0, Q, y, P ;$
        \STATE
            Compute the Jacobian $J$ of $g$;
        \STATE
            Solve linear system \eqref{eqn: Newton} to obtain $\Delta y$ and $\Delta \lambda ;$
        \STATE
            $y \gets y + \Delta y ;$
        \STATE
            $\lambda \gets \lambda + \Delta \lambda ;$
        \IF{$ \| g(y) \| < \sigma_1$ and $ \| \nabla L \| < \sigma_2$}
        \STATE
            \textbf{break}
        \ENDIF
            
        \ENDFOR
        
    \end{algorithmic}
\end{algorithm}

\section{Experimental Results}
\begin{figure}[t]
    \centering
    \includegraphics[width=1\textwidth]{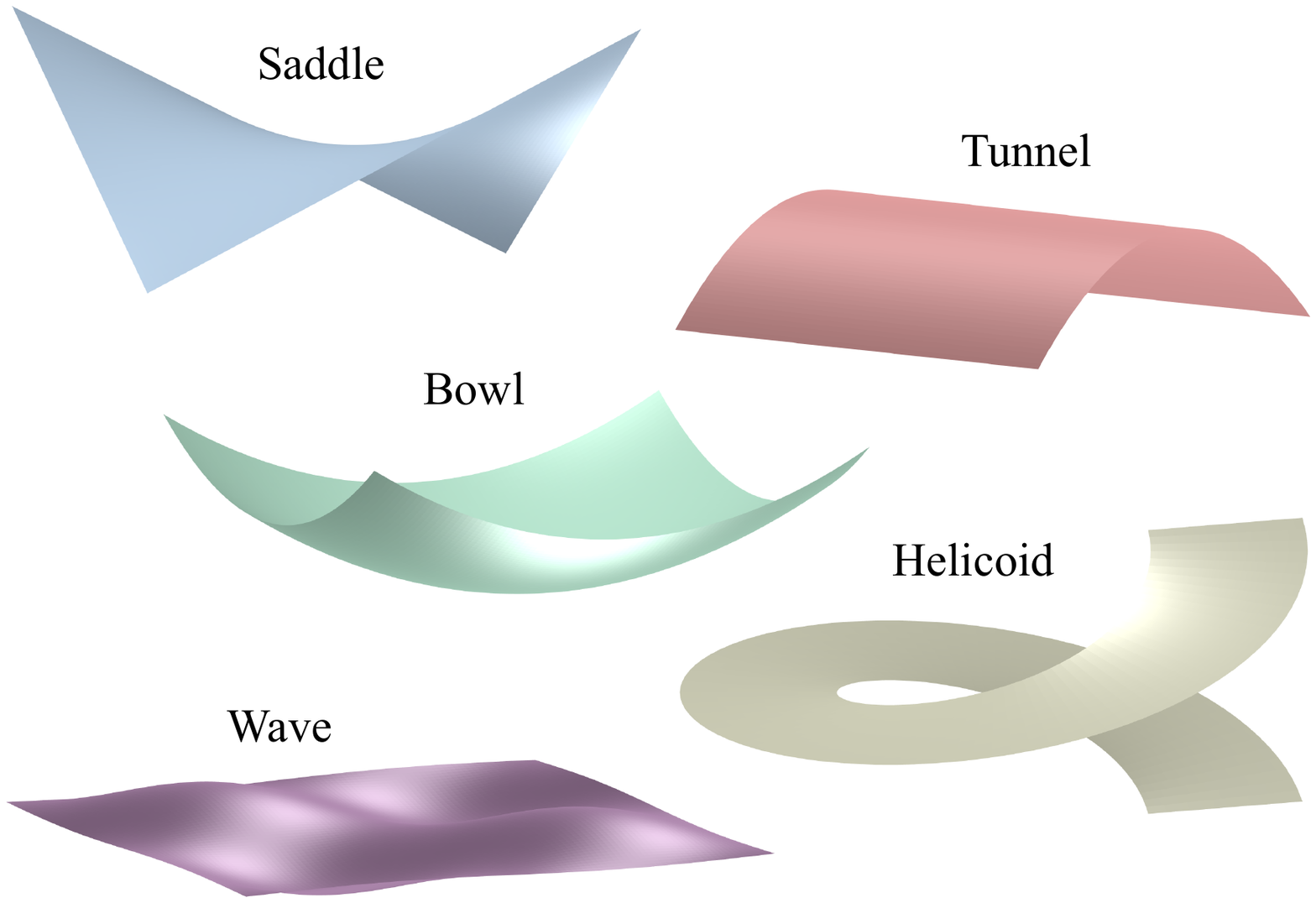}
    \caption{
        Different target surfaces considered in our experiments: tunnel, saddle, bowl, helicoid, and wave. 
    }
    \label{fig: surfaces}
\end{figure}

With the framework established for optimization and initial guess generation, we conducted experiments to investigate the importance of energy terms, and the flexibility and efficacy of our methods are also examined by testing a variety of surfaces. 
The feasibility of our methods applied to different structures of the Miura-ori pattern is investigated with quantitative performance. 
In Figure \ref{fig: surfaces}, five surfaces, including saddle, tunnel, bowl, wave and helicoid are illustrated, as the test surfaces in the experiments. 

\subsection{Ablation Study}
To address the significance of each energy term in the optimization framework, we conducted experiments by excluding one energy term at a time while retaining the other two. Saddle surface, with varying sizes, is used as input surface throughout the ablation. This approach allowed us to isolate and observe the contribution of each term to the overall performance of the Miura-ori pattern design. To better observe the results, the experiments were conducted with an early stopping mechanism. Without early stopping, the phenomena would be even more exaggerated. 

\begin{figure}[t]
    \centering
    \includegraphics[width=\textwidth]{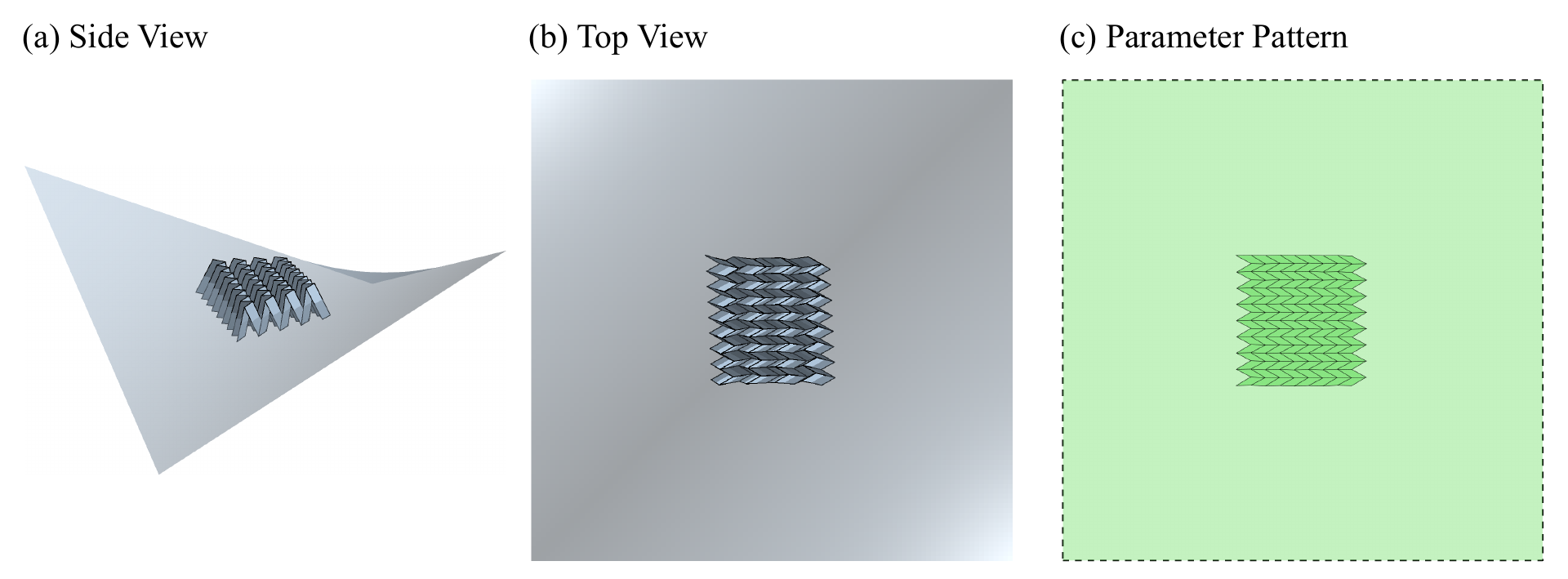}
    \caption{
        Shrinkage effect. (a) and (b) show the folded pattern in two views. (c) shows the shrunken parameter pattern in the parameter domain. 
    }
    \label{fig: ablation: shrinkage}
\end{figure}

Firstly, the edge length energy in Eq.~\eqref{ene: edge length} was dropped, which serves to regularize the geometry of the folded pattern, with Figure \ref{fig: ablation: shrinkage} illustrating the result. In the absence of the edge length regularizer, the folded pattern on the saddle surface demonstrated noticeable shrinkage at its center, a region that is approximately flat geometry where the feasibility constraints can be easily satisfied. As a result, the designed pattern fails to approximate the surface geometry. Without early stopping, the pattern would continue to shrink excessively. These observations highlight the critical role of the edge length energy in maintaining the geometry of the folded pattern. 

\begin{figure}[t]
    \centering
    \includegraphics[width=\textwidth]{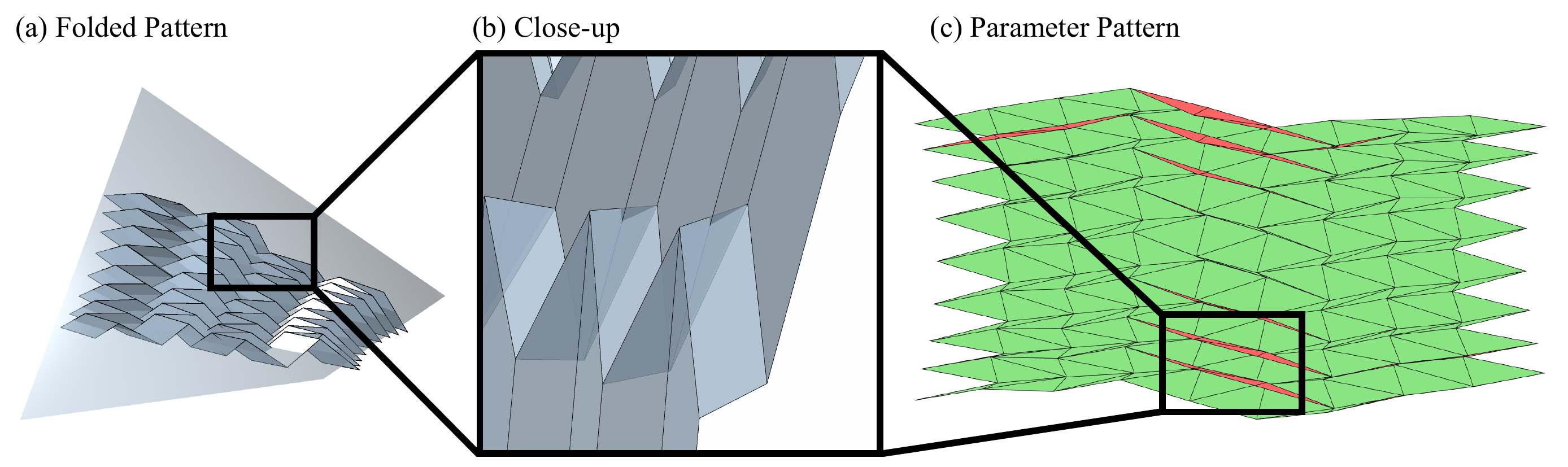}
    \caption{
        Self-intersection effect. (a) shows a result that drops Beltrami energy. (b) focuses on a region where self-intersection is observed. (c) shows the Beltrami coefficient of the resulting parameter pattern, where triangles with $|\mu| \ge 1$ are drawn in red. 
    }
    \label{fig: ablation: self-intersection}
\end{figure}

Next, the effect of excluding the quasiconformality energy in Eq.~\eqref{ene: beltrami}, which regularizes the parameter pattern, was investigated, with results illustrated in Figure \ref{fig: ablation: self-intersection}. 
The figure shows significant deformation in the parameter pattern. With no controls on deformation, certain regions, which are drawn in red in the figure, attain a Beltrami coefficient with magnitude exceeding 1. Such overlaps may result in self-intersection of the folded pattern, which is physically impossible as a Miura-ori pattern. A close-up of the phenomenon is shown in Figure \ref{fig: ablation: self-intersection}(b). 

\begin{figure}[t]
    \centering
    \includegraphics[width=\textwidth]{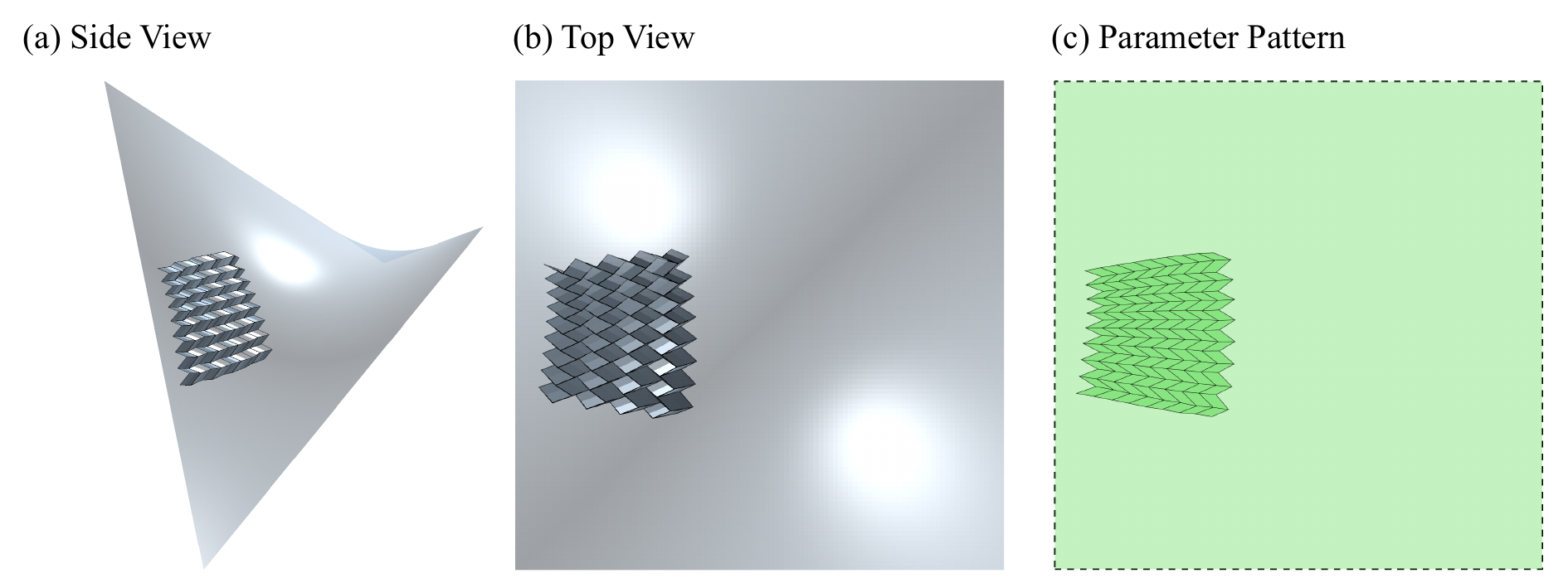}
    \caption{
        Off-centering effect. (a) and (b) shows the folded pattern in two views. (c) shows the skewed parameter pattern in the extended parameter domain. 
    }
    \label{fig: ablation: off-centering}
\end{figure}

Finally, the impact of excluding the centering energy term in Eq.~\eqref{ene: centering} was examined, with results illustrated in Figure \ref{fig: ablation: off-centering}. In this experiment, the surface and parameter domain were extended while the size of the initial guess remained unchanged, in order to exaggerate the effect. 
In the figure, a clear off-centering effect was observed in both the resulting folded and parameter patterns. The outer region of the surface is approximately linear, where the feasibility could be easily satisfied. 
This observation suggests the importance of the centering energy in maintaining the patterns in a desired position.

\subsection{Surface-Specific Performance}
\begin{figure}[t!]
    \centering
    \includegraphics[width=\textwidth]{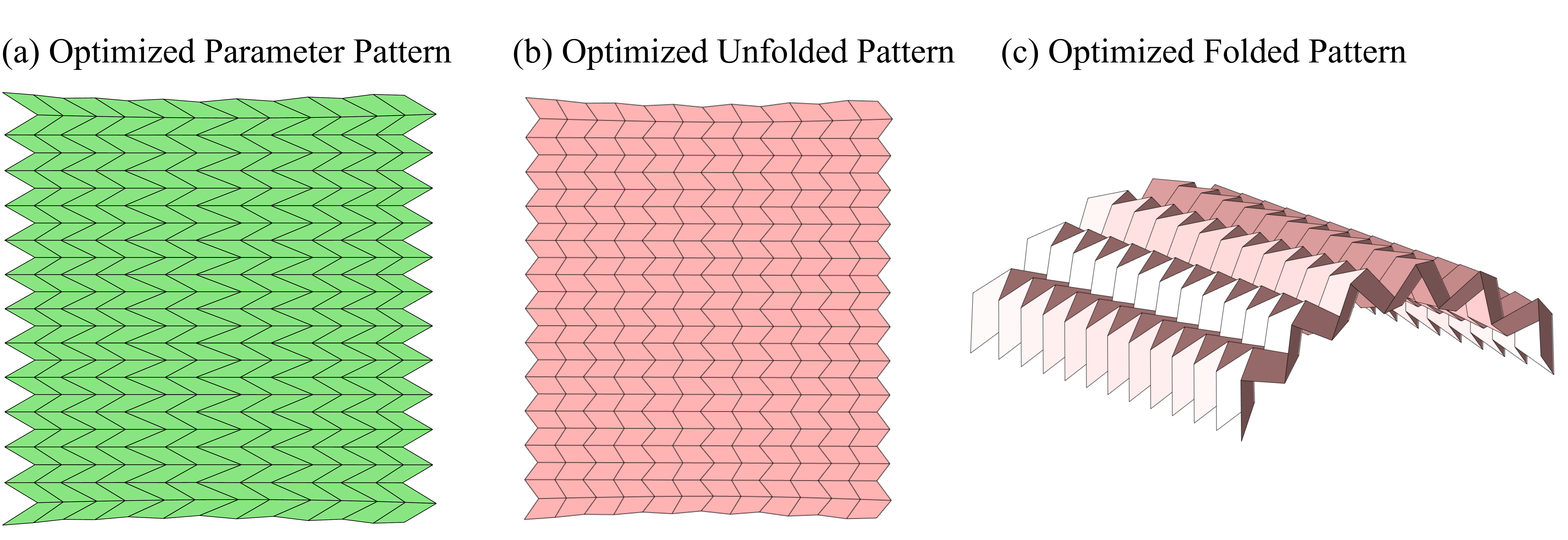}
    \caption{
        Experimental results for tunnel. (a) is the optimized planar pattern in the parameter domain. (b) is the unfolded pattern of the optimized folded pattern, and (c) is the optimized folded pattern.
    }
    \label{fig: exp tunnel}
\end{figure}

\begin{figure}[t!]
    \centering
    \includegraphics[width=1\textwidth]{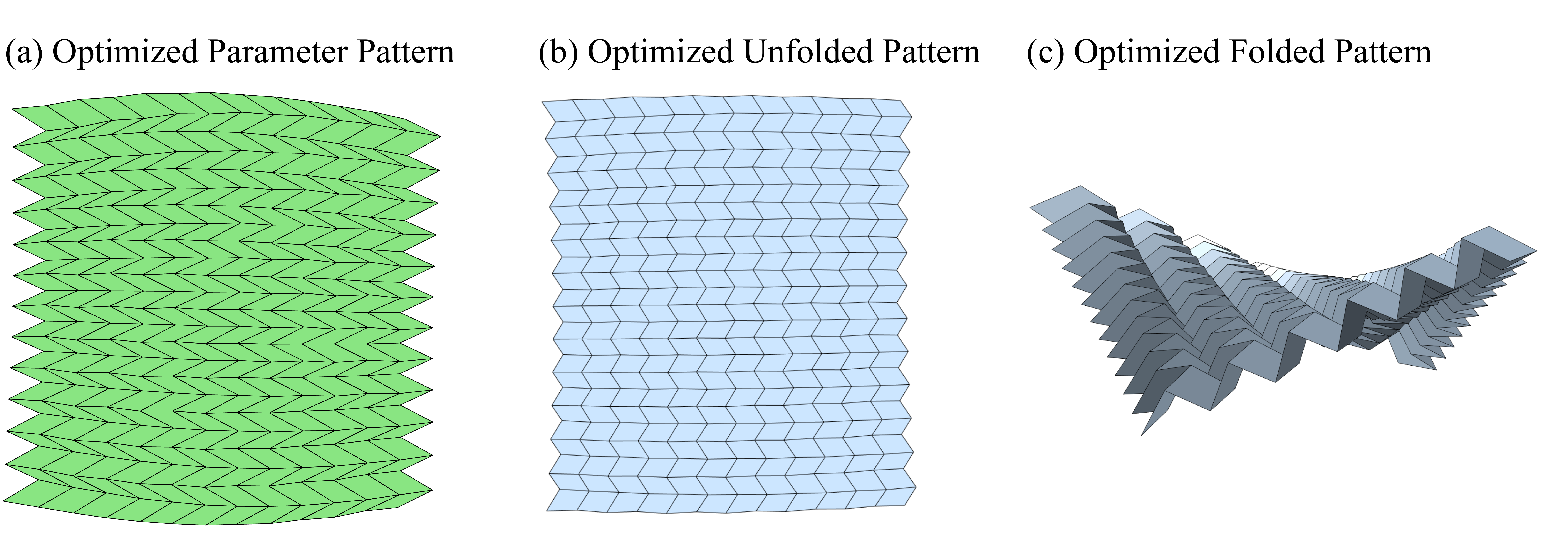}
    \caption{
        Experimental results for saddle. 
        (a) is the optimized planar pattern in the parameter domain. (b) is the unfolded pattern of the optimized folded pattern, and (c) is the optimized folded pattern.
    }
    \label{fig: exp saddle}
\end{figure}

\begin{figure}[t!]
    \centering
    \includegraphics[width=1\textwidth]{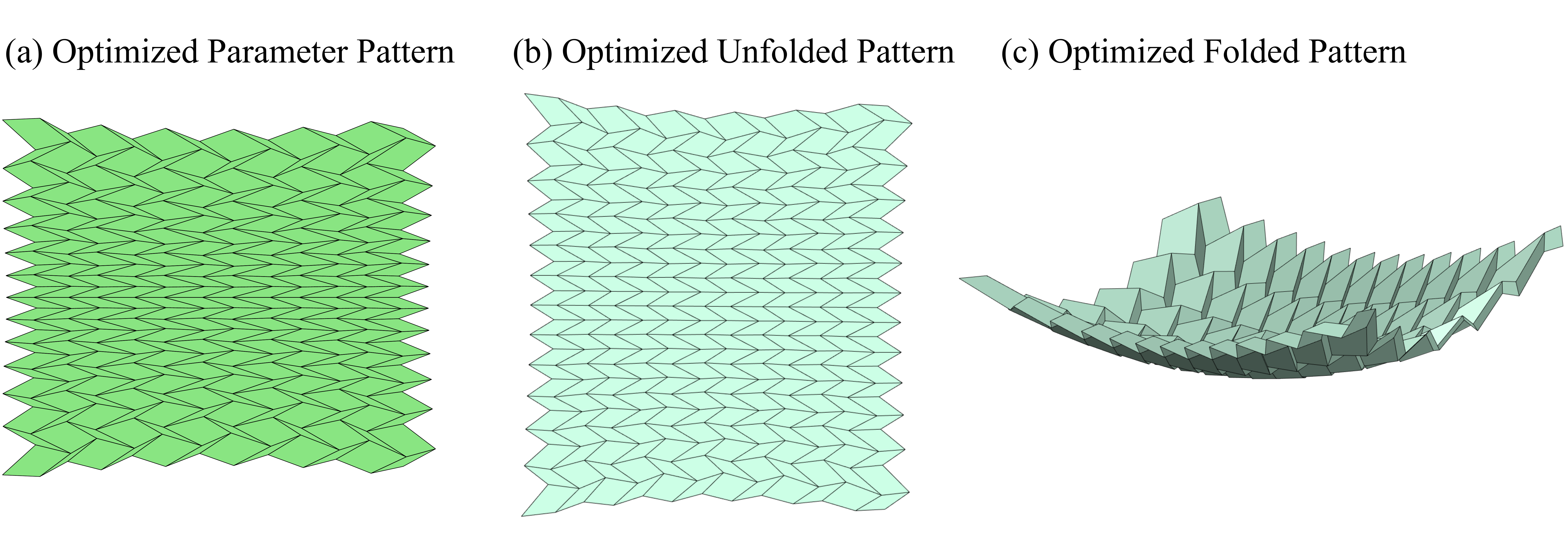}
    \caption{
        Experimental results for bowl. 
        (a) is the optimized planar pattern in the parameter domain. (b) is the unfolded pattern of the optimized folded pattern, and (c) is the optimized folded pattern.
    }
    \label{fig: exp bowl}
\end{figure}

\begin{figure}[t!]
    \centering
    \includegraphics[width=1\textwidth]{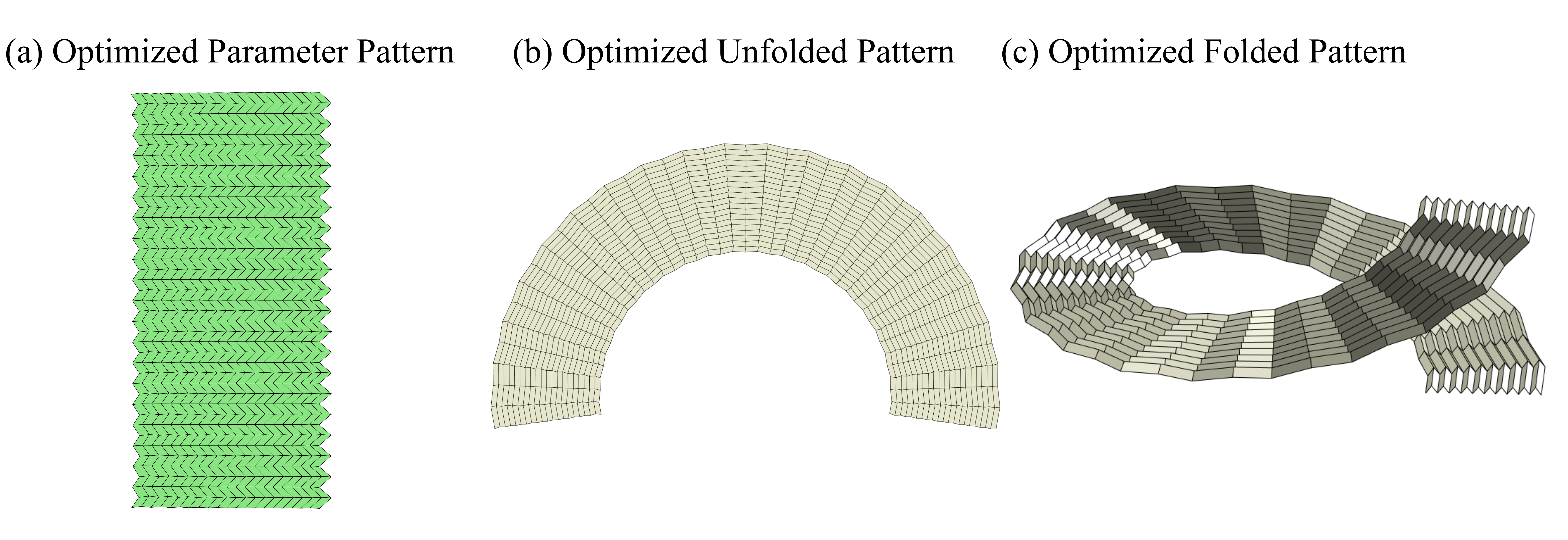}
    \caption{
        Experimental results for helicoid. 
        (a) is the optimized planar pattern in the parameter domain. (b) is the unfolded pattern of the optimized folded pattern, and (c) is the optimized folded pattern.
    }
    \label{fig: exp helicoid2}
\end{figure}

\begin{figure}[t!]
    \centering
    \includegraphics[width=1\textwidth]{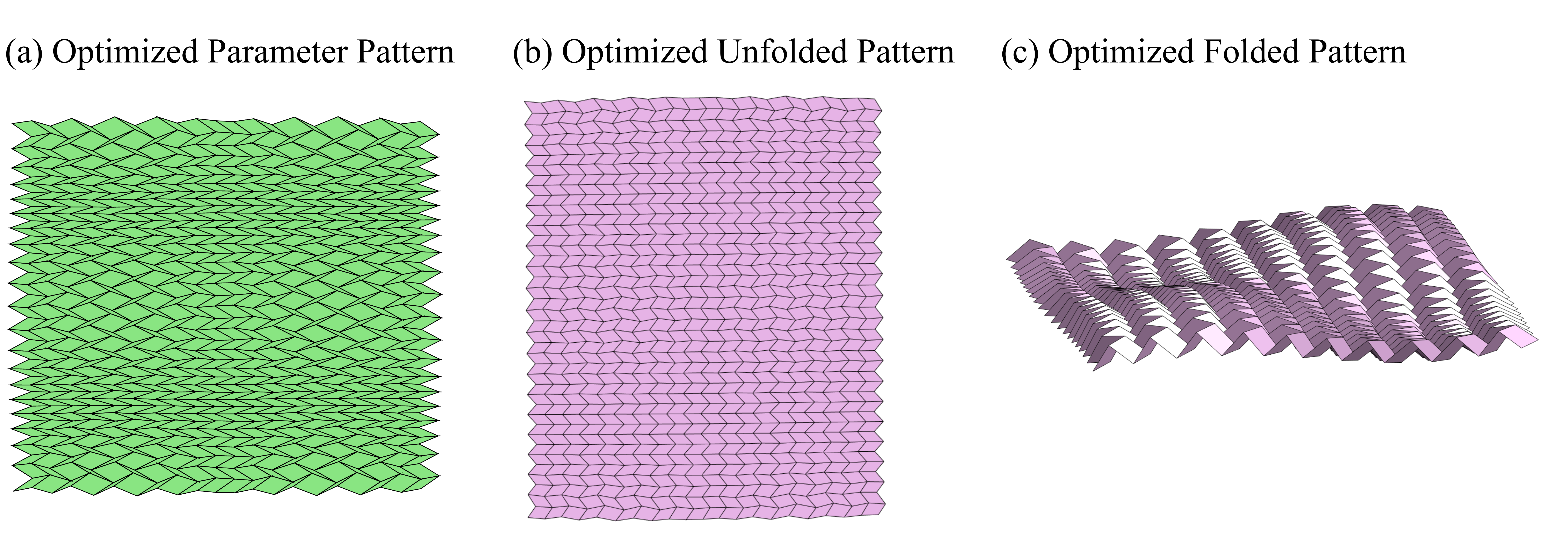}
    \caption{
        Experimental results for wave. 
        (a) is the optimized planar pattern in the parameter domain. (b) is the unfolded pattern of the optimized folded pattern, and (c) is the optimized folded pattern.
    }
    \label{fig: exp wave}
\end{figure}

To evaluate the flexibility of the proposed optimization framework, the detailed performance on the five surfaces illustrated in Figure \ref{fig: surfaces} are examined. 
The results are presented in Figures \ref{fig: exp tunnel} for tunnel, \ref{fig: exp saddle} for saddle, \ref{fig: exp bowl} for bowl, \ref{fig: exp helicoid2} for helicoid, and \ref{fig: exp wave} for wave. Each figure illustrates the optimized parameter pattern and folded pattern, together with the corresponding unfolded pattern.

Following the optimization process, all the folded patterns satisfied the planarity and developability constraints, with minimal error due to the limitations of numerical computation. 
The maximum errors observed for both planarity and developability across the tested surfaces are presented in Table \ref{tab: feasibility error and deformation}. In our computations, planarity is quantified as the volume of the parallelepiped formed by the four nodes of each quadrilateral face, while developability is measured as the difference between the sum of the four angles around each interior node and $2\pi$. The errors reported in Table \ref{tab: feasibility error and deformation} are significantly small. Furthermore, the resulting unfolded patterns exhibit a high degree of flatness, confirming the successful construction of valid generalized Miura-ori patterns. This flatness, combined with the low error margins, validates the effectiveness of our approach in generating practical origami designs.

\begin{table}[t!]
    \centering
    \begin{tabular}{ |p{1.5cm}||p{2.3cm}|p{2.3cm}||p{2.3cm}|p{2.3cm}| }
        \hline
        & \multicolumn{2}{|c||}{Feasibility Error} & \multicolumn{2}{c|}{Regularity} \\
        \hline
        Surface & $\max |g_{\text{planarity}}|$ & $\max |g_{\text{develop}}|$ & $E_l$ & mean $ |\mu|$ \\
        \hline
        Tunnel   & $8.35 \times 10^{-18}$ & $8.89 \times 10^{-15}$ & $5.51 \times 10 ^ {-4}$ & $0.0819$ \\
        Saddle   & $1.27 \times 10^{-18}$ & $2.66 \times 10^{-15}$ & $5.09 \times 10 ^ {-5}$ & $0.0952$ \\
        Bowl     & $2.66 \times 10^{-18}$ & $3.55 \times 10^{-15}$ & $9.29 \times 10 ^ {-4}$ & $0.277$ \\
        Helicoid & $6.80 \times 10^{-16}$ & $7.11 \times 10^{-15}$ & $1.43 \times 10 ^ {-4}$ & $0.0762$ \\
        Wave     & $3.12 \times 10^{-16}$ & $1.07 \times 10^{-14}$& $6.09 \times 10 ^ {-4}$ & $0.231$ \\
        \hline
    \end{tabular}
    \caption{Table for feasibility error and measurements for deformation. }
    \label{tab: feasibility error and deformation}
\end{table}

In Table \ref{tab: feasibility error and deformation}, the average magnitudes of the Beltrami coefficients and final values of edge length energy $E_l$ are also included. 
$E_l$ of all surfaces are kept in low values, indicating that the size of folded patterns is well preserved, and hence the desired geometric features. 
For tunnel, saddle and helicoid surfaces, the parameter patterns before and after optimization exhibit notable similarity, as shown in Figure \ref{fig: exp tunnel}, \ref{fig: exp saddle} and \ref{fig: exp helicoid2}. This corresponds to their relatively low Beltrami coefficients, indicating less geometric distortion. 
In contrast, the bowl and wave surfaces display more pronounced deformations in the parameter pattern after optimization in Figure \ref{fig: exp bowl} and \ref{fig: exp wave}. This is consistent with their higher Beltrami coefficient.

\subsection{Structure Parameter Analysis}

In our methods, there are two pre-defined factors, thickness and resolution, which directly affect the structure and performance of the results. In this section, we examine the effect of these variables on the saddle surface with their quantitative performance. 

\begin{figure}[t]
    \centering
    \includegraphics[width=1\textwidth]{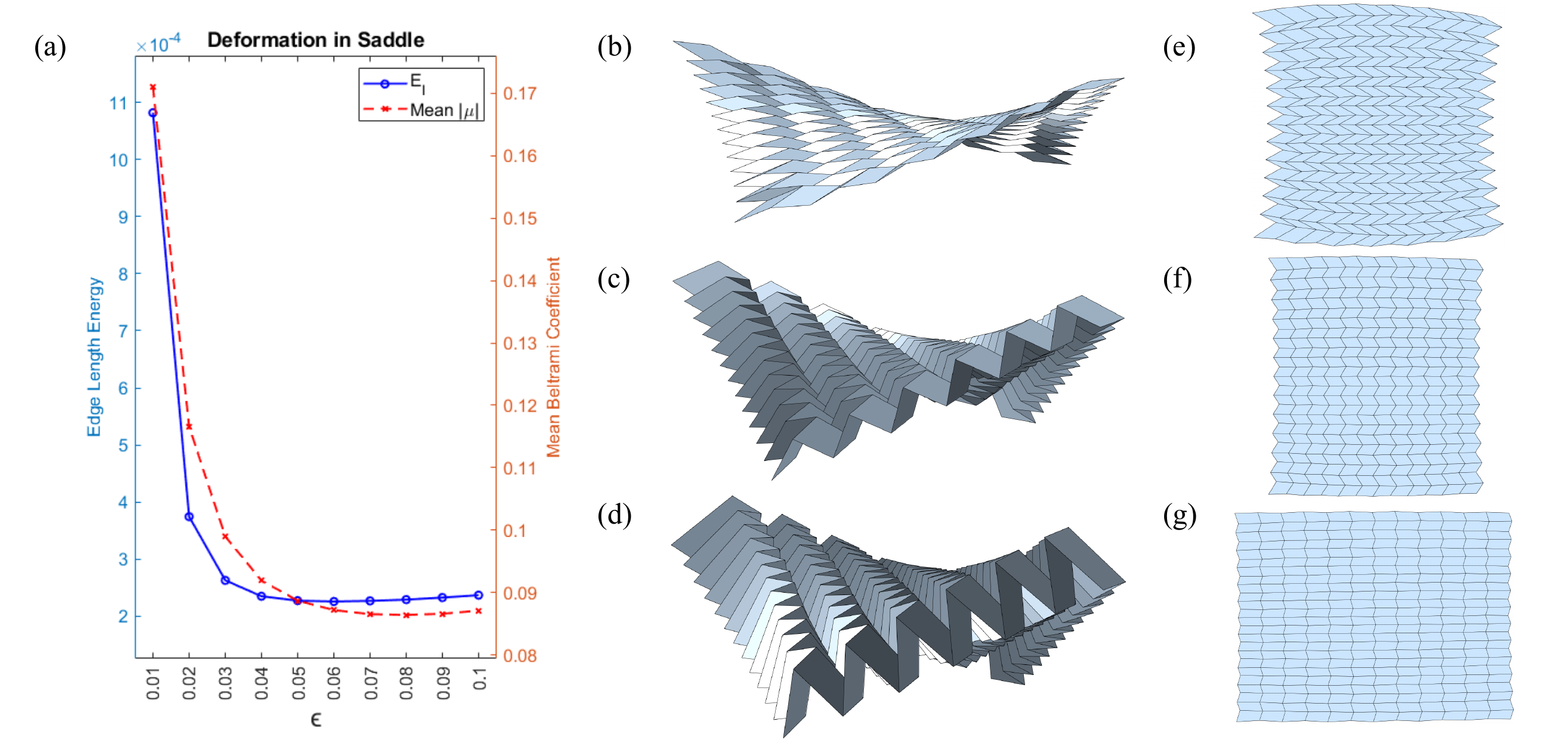}
    \caption{
        Experimental results for the effects of different thickness of folded patterns on the saddle surface. 
    }
    \label{fig: epsilon saddle}
\end{figure}

First, the effect of various thickness of folded pattern is examined, specifically $\epsilon = 0.01, 0.02, 0.03, \dots, 0.1$ are chosen for the model. 
The experimental results are summarized in Figure \ref{fig: epsilon saddle}. 
In the figure, (a) shows the effects on the edge length energy and the average norm of the Beltrami coefficient (mean $|\mu|$) by changing the size of $\epsilon$. (b), (c) and (d) in the middle column highlight the optimized folded pattern with $\epsilon = 0.01, 0.05$ and $0.1$. (e), (f) and (g) in the last column show the corresponding unfolded patterns. All of the experiments satisfy the planarity constraint and the developability constraint with a numerical error in the order of $10^{-17}$ and $10^{-14}$ respectively. 
As shown in Figure \ref{fig: epsilon saddle}, increasing $\epsilon$ reduces the overall deformation at first, while further increasing it does not drastically change the deformation. Despite giving less deformation, a large value in $\epsilon$ may give a less desirable result. A smaller thickness, as shown in (b), allows a folded pattern that better approximates the saddle surface. Meanwhile, a thicker pattern, as in (d), the folded pattern deviates more from the input surface. A trade-off between surface alignment and deformation can be observed. 

\begin{figure}[t]
    \centering
    \includegraphics[width=1\textwidth]{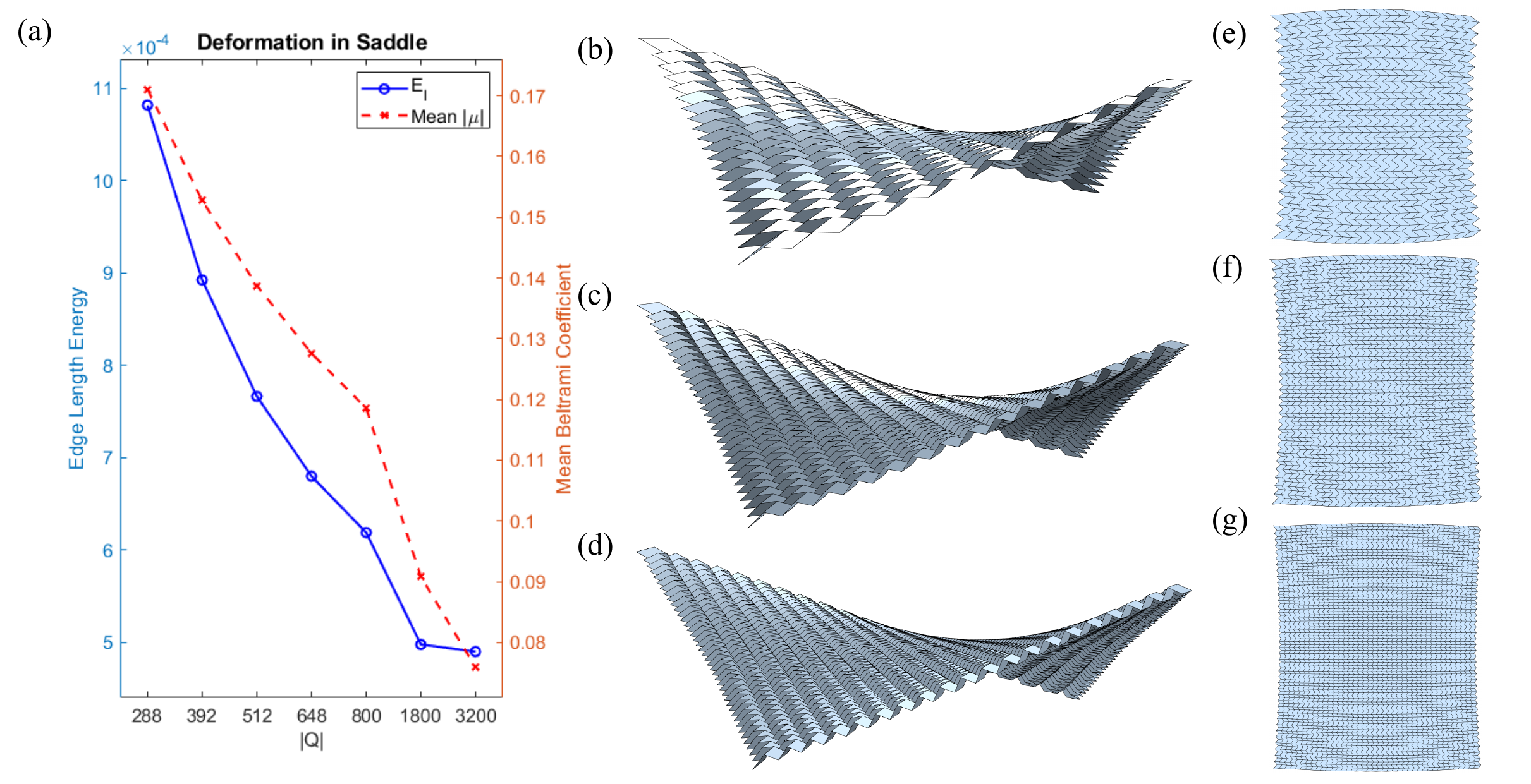}
    \caption{
        Experimental results for the effects of different resolutions of folded patterns on the saddle surface. 
    }
    \label{fig: saddle resolution}
\end{figure}

\begin{figure}[t]
    \centering
    \includegraphics[width=1\textwidth]{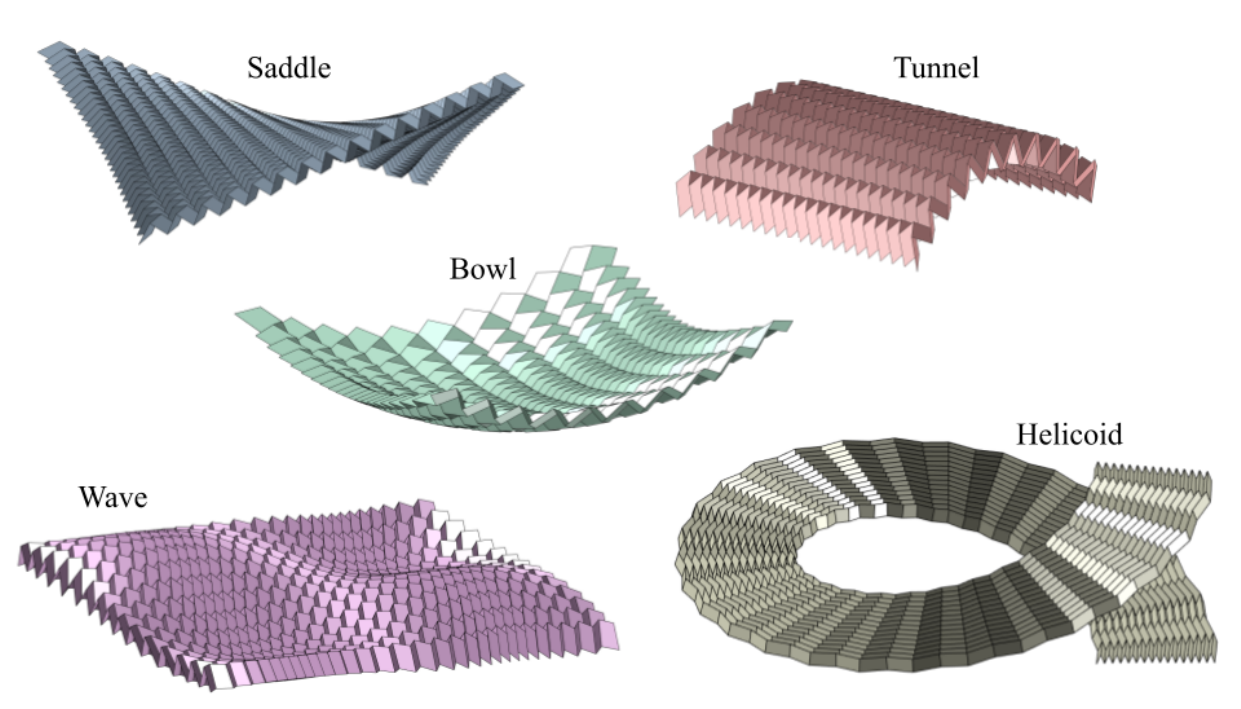}
    \caption{
        Experimental results for saddle, tunnel, bowl, helicoid, and wave with higher resolution.
    }
    \label{fig: exp dense}
\end{figure}

On the other hand, in Figure \ref{fig: saddle resolution}, experiments are conducted, with fixed $\epsilon$ and different numbers of faces, including $|Q| = 288, 392, 512, 648, 800, 1800, 3200$. In the figure, (a) shows the effects of resolution to the edge length energy and mean $|\mu|$, (b), (c) and (d) shows the folded optimized patterns with $|Q| = 800, 1800$, and $3200$ respectively, and (e), (f) and (g) shows the corresponding unfolded patterns. 
Without increasing $\epsilon$ like the previous experiments, the deformation reduces as the resolution increases, since the flexibility increases. 
Experiments with higher resolution for all the five surfaces are conducted, and the resulting folded patterns are shown in Figure \ref{fig: exp dense}, all achieving almost zero unfoldability error, and low deformation. 
Although increasing the resolution gives better result, a large number of variables comes at a cost. First, the computational cost increases, as the optimization model involves the inverse of large matrices. Also, the difficulty of folding the pattern physically rises as more creases are involved, and the error of misfolding would accumulate throughout the pattern.

\section{Conclusions and future directions}
\label{sec: conclusions}

In this study, we have formulated the surface-aligned Miura-ori inverse design problem as a mapping problem and addressed it through constrained optimization. Provided a parameterized input surface, upper and lower surfaces are constructed, with an initial folded pattern embedded between them. By this construction, we can ensure the folded pattern aligns the surface, with a pre-designed discrepancy. The folded pattern is then parameterized into the common planar domain with the surface, and optimized under specific unfoldability constraints and shape regularizers. Quasiconformality is included as an optimization objective. It can ensure the bijectivity of mappings in the parameter domain, which empirical experiments suggest that it can avoid a potential problem of self-intersection in the folded pattern.  
Apart from the quasiconformality, an edge length energy and a centering energy are included for further regularizing the result, and ablation studies are performed to address the significance of each energy term. Also, in order to demonstrate the flexibility of our methods, several surfaces are tested, all with promising results. The various values in structure factors, including thickness and resolution, are considered. By increasing the thickness, the difficulty in the optimization decreases, but the discrepancy to the input surface also increases. Similarly, a higher resolution reduces optimization difficulties, but a higher computational cost is required. 

Looking ahead, our methods can be extended to explore surfaces with more complex geometries and topologies. Mathematical and physical properties of the designed structures can be analyzed, and more geometric or mechanical regularizers can also be incorporated into the model. The optimization strategy can be refined, potentially using neural networks, to improve the optimization quality. Extending our proposed framework toward kirigami can also be considered.

\section*{Acknowledgement} Lok Ming Lui is supported by HKRGC GRF (Project ID: 14306723).

\bibliographystyle{siamplain}
\bibliography{references}

\end{document}